\documentclass[10pt]{article}
\usepackage[top=2cm, left=2cm, right=2cm, bottom=2cm]{geometry}
\usepackage{import}
\usepackage{url}
\usepackage{graphicx}
\usepackage{amsmath, amssymb, amsfonts, amsthm}
\usepackage{tikz} 

\newcounter{row}
\newcounter{col}

\newcommand\setrow[4]{
  \setcounter{col}{1}
  \foreach \n in {#1, #2, #3, #4} {
    \edef\x{\value{col} - 0.5}
    \edef\y{4.5 - \value{row}}
    \node[anchor=center] at (\x, \y) {\n};
    \stepcounter{col}
  }
  \stepcounter{row}
}

\newcommand{\detY}{
\begin{tikzpicture}[scale=.2]
  \begin{scope}
    \draw (0, 0) grid (4, 4);

    \setcounter{row}{1}
    \setrow {$\bullet$}{$\bullet$}{$\bullet$}{$\bullet$}  
    \setrow {$\bullet$}{$\bullet$}{$\bullet$}{$\bullet$}  
    \setrow {$\bullet$}{$\bullet$}{$\bullet$}{$\bullet$}  
    \setrow {$\bullet$}{$\bullet$}{$\bullet$}{$\bullet$}

  \end{scope}

\end{tikzpicture}
}

\newcommand{\detI}{
\begin{tikzpicture}[scale=.2]
  \begin{scope}
    \draw (0, 0) grid (4, 4);

    \setcounter{row}{1}
    \setrow {.}{.}{.}{.}  
    \setrow {.}{.}{.}{.}  
    \setrow {.}{.}{.}{.}  
    \setrow {.}{.}{.}{.}

  \end{scope}

\end{tikzpicture}
}

\newcommand{\detZC}{
\begin{tikzpicture}[scale=.2]
  \begin{scope}
    \draw (0, 0) grid (4, 4);

    \setcounter{row}{1}
    \setrow { }{ }{$\bullet$}{ }  
    \setrow { }{ }{$\bullet$}{ }  
    \setrow { }{ }{$\bullet$}{ }  
    \setrow { }{ }{$\bullet$}{ }

  \end{scope}

\end{tikzpicture}
}

\newcommand{\detCZ}{
\begin{tikzpicture}[scale=.2]
  \begin{scope}
    \draw (0, 0) grid (4, 4);

    \setcounter{row}{1}
    \setrow { }{ }{ }{ }  
    \setrow { }{ }{ }{ }  
    \setrow {$\bullet$}{$\bullet$}{$\bullet$}{$\bullet$}  
    \setrow { }{ }{ }{ }

  \end{scope}
\end{tikzpicture}
}

\newcommand{\detCZC}{
\begin{tikzpicture}[scale=.2]
  \begin{scope}
    \draw (0, 0) grid (4, 4);

    \setcounter{row}{1}
    \setrow { }{ }{ }{ }  
    \setrow { }{ }{ }{ }  
    \setrow { }{ }{$\bullet$}{ }  
    \setrow { }{ }{ }{ }

  \end{scope}

\end{tikzpicture}
}

\newcommand{\detZCZ}{
\begin{tikzpicture}[scale=.2]
  \begin{scope}
    \draw (0, 0) grid (4, 4);

    \setcounter{row}{1}
    \setrow {$\bullet$}{$\bullet$ }{.}{$\bullet$ }  
    \setrow {$\bullet$ }{$\bullet$ }{.}{$\bullet$ }  
    \setrow {.}{.}{.}{.}  
    \setrow {$\bullet$ }{$\bullet$ }{.}{$\bullet$ }

  \end{scope}

\end{tikzpicture}
}
\title{Unitarization of Pseudo-Unitary Quantum Circuits in the S-matrix Framework}

\author{Dennis Lima$^{1}$ and Saif Al-Kuwari$^{1}$
\\$^1$ Qatar Center for Quantum Computing (QC2), Qatar
\\deaq54989@hbku.edu.qa and smalkuwari@hbku.edu.qa
}
\date{\today}

\begin{document}

\maketitle

\begin{abstract}
Pseudo-unitary circuits are recurring in both S-matrix theory and analysis of No-Go theorems.
We propose a matrix and diagrammatic representation for the operation that maps S-matrices to T-matrices and, consequently, a unitary group to a pseudo-unitary one. We call this operation ``partial inversion'' and show its  diagrammatic representation in terms of permutations. We find the expressions for the deformed metrics and deformed dot products that preserve physical constraints after partial inversion.
Subsequently, we define a special set that allows for the simplification of expressions containing infinities in matrix inversion. Finally, we propose a renormalized-growth algorithm for the T-matrix as a possible application. 
The outcomes of our study expand the methodological toolbox needed to build a family of pseudo-unitary and inter-pseudo-unitary circuits with full diagrammatic representation in three dimensions, so that they can be used to exploit pseudo-unitary flexibilization of unitary No-Go Theorems and renormalized circuits of large scattering lattices.
\end{abstract}

%
\vspace{2pc}

\section{\label{sec1}Introduction}
    Over the years, pseudo-unitary quantum mechanics has rapidly grown from the interest in  pseudo-Hermitian Hamiltonians \cite{mostafazadeh2004pseudounitary} in early 2000 to describe PT-symmetric quantum systems \cite{mostafazadeh2003exact}, and recently to simulate and experiment pseudo-unitary quantum cloning and quantum deletion in optical circuits \cite{chen2021quantum,zhan2020experimental}. Furthermore, controlling Hilbert-space inner products \cite{karuvade2022observing} for fast quantum algorithms in PT-symmetric systems \cite{bender2007faster,bender2002complex} is a direct implication of knowing how to manipulate pseudo-unitary groups. A consistent and experiment-friendly explanation of how pseudo-unitary algebras are built based on transformations of Hilbert spaces is given by Karuvade \emph{et al.} \cite{karuvade2022observing}.

    Another motivation to study pseudo-unitary circuits comes from the S-matrix theory. The general solution $\psi(x)$ to the Schrödinger equation for a one-dimensional wave scattering across a localized potential is equation (\ref{sol}), with $\psi_{ab}(x)$ the solution between $x=a$ and $x=b$, the region where the potential is non-zero.
    \begin{align} \label{sol}
        \psi(x) = \begin{cases}
            \psi_{00} e^{ikx} + \psi_{01} e^{-ikx},         &\quad x < a;
            \\  \psi_{ab}(x), &\quad a<x<b;
            \\  \psi_{10} e^{ikx} + \psi_{11} e^{-ikx}
            &\quad x>b.
        \end{cases}
    \end{align}

    \begin{align} \label{Smat}
        \begin{pmatrix}
            \psi_{01} \\ \psi_{10}
        \end{pmatrix}
        =S
        \begin{pmatrix}
            \psi_{00} \\ \psi_{11}
        \end{pmatrix}.
    \end{align}

    If the S-matrix of a system is unitary (equation (\ref{sss})), then the transfer matrix (equation (\ref{ttt})) derived from it is known to be pseudo-unitary with respect to either the $\sigma_3 \otimes I_{m}$ metric  \cite{beenakker1997random} or the $I_m \otimes \sigma_3$ metric \cite{snyman2008calculation}, where $\sigma_3=|0 \rangle \langle 0| - |1 \rangle \langle 1|$ is a Pauli matrix and $I_{m}$ is the identity matrix with dimensions $(m,m)$. In equations (\ref{sss}) and (\ref{ttt}), $t$, $t^\prime$, $r$ and $r^\prime$ are Hermitian matrices \cite{beenakker1997random}.
    \begin{align} \label{sss}
    S &= \begin{pmatrix} t & r \\ -r^\prime & t^\prime
    \end{pmatrix}, 
    \quad \text{with} \quad tt^\prime + rr^\prime = I_m;
    \\ T &=
    \begin{pmatrix}
        T^{++} & T^{+-}
        \\ T^{-+} & T^{--}
    \end{pmatrix}
    =
    \begin{pmatrix}
    (t^\prime)^{-1} &     r(t^\prime)^{-1}
   \\     (t^\prime)^{-1}r^\prime &  (t^\prime)^{-1}
    \end{pmatrix}.
    \label{ttt}
    \end{align}

    The S-matrix theory was initially proposed by Wheeler \cite{wheeler1937mathematical} to study subatomic particles. It provides maximal analyticity, but cannot compute interparticle forces \cite{chew1964search}. In many-body systems of electrons, the S-matrix and, more often, the T-matrix, are alternative formalisms to analyse critical behaviour, conductance and localization in scattering processes \cite{markos2008wave}.

\begin{figure}[h]
    \centering
    	\caption{S-matrix and T-matrix scattering diagrams for two-component spinors. \label{scatd}}
	    \includegraphics[width= 0.5 \textwidth]{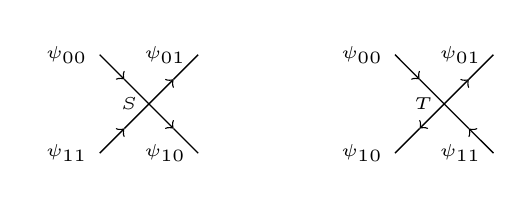}
    \end{figure}

    Feynman diagrams \cite{feynman1949space} or, equivalently, quantum circuits, provide us with a visual interpretation of both S and T matrices by rearranging the arrows that represent states in a scattering process (figure \ref{scatd}) \cite{chalker1988percolation}. Lattices are grown by juxtaposing these diagrams side by side. This ease of growing periodic lattices with T-matrices may motivate the preference to use T-matrix formalism for periodic potentials \cite{alhassid1983potential}.
    
    Recent studies towards experimental control of pseudo-unitary operators focus on PT-symmetric systems, but lack mentions of examples from the T-matrix theory \cite{chen2021quantum,karuvade2022observing}. 
    As consequence, the pseudo-unitarity of the T-matrix formalism is restricted to analytical studies for calculation of conductance, specially in conductors like graphene \cite{snyman2008calculation}. 
    
    The design of inter-unitary protocols has the advantage to allow for the use of a matrix product in place of a tensor product for calculation of physical properties like the Landauer conductance in equation (\ref{landauer}). This possibility is clarified by the diagrammatic representation of scattering lattices and the change of direction of diagrammatic growth after solving a linear system of equations \cite{snyman2008calculation}.

\begin{align} \label{landauer}
    G = \frac{4e^2}{h} \mathrm{Tr} 
    \left\{
    [ (T^{--})^{\dagger} T^{--} ]^{-1} 
    \right\}
\end{align}

    Although such protocols that take advantage of matrix products in different pseudo-unitary groups have been used \cite{snyman2008calculation}, the formal transformation that allows it and its mathematical properties in the scope of pseudo-unitary groups are not well understood.

    We extend it and provide tools to build inter-unitary circuits, i.e., circuits that make use of more than one pseudo-unitary group, by defining the matrix representation for the transformation algorithm that performs a step-wise matrix inversion necessary to convert between S-matrix and T-matrix formalisms. We show that it contains a representation of the symmetric group, can be used to transit between pseudo-unitary groups, leads to the definition of a special set useful to manipulate singular matrices and is equipped with the diagrammatic representation to further conceive a renormalized quantum circuit growth.

\subsection{Contributions}
The contributions of this paper can be summarized as follows: 
\begin{enumerate}
    \item proposition of partial inversion algorithm with an operator and diagrammatic representations; 
    \item proposition of a heuristic diagram that maps a family of pseudo-unitary groups to one another through partial inversion; 
    \item definition of the effect of partial inversion on dot product metric with conservation of dot product; 
    \item demonstration of the application of partial inversion to invert and invert back matrices with zeros; 
    \item demonstration of the application of partial inversion to grow the dimensions of a circuit, while the dimension of the matrix representing the circuit is kept constant.
\end{enumerate}

\subsection{Organization}
The rest of the paper is organized as follows: 
    In section \ref{sec3}, we motivate the study of partial inversion in S-matrix theory with the effect of Legendre transform in dot products. Subsequently, in section \ref{sec4}, we reformulate partial inversion as an operator in terms of matrices and propose a diagrammatic representation that is consistent with representations of the symmetric group. We follow to defining heuristically the diagram that solves the deformed algebra equation of non-commutativity of partial inversion and we show that it maps a family of $SU(2,\eta)$ groups to one another. Next, we present a unitary representation of a sequence of partial inversions and we explore how partial inversion deforms the dot product by deforming the conservation metric $g$. 
    Finally, we discuss two potential applications in section \ref{sec5}. The first one is the use of partial inversion to invert matrices forth and back even in the presence of inverted zeros (non-finite matrices). The second one is the use of partial inversion as a means to extend unitary circuits in two dimensions without changing the dimensions of the matrix that represents the circuit. We call this approach "renormalized growth". We conclude the paper in section \ref{sec6} with a few final remarks.

\section{\label{sec3}Methods}

In this section we furnish the dot product conservation equation in a set of Legendre-transformed functions as a motivational model. Subsequently, in the light of the Gauss-Jordan elimination algorithm, we introduce a typical linear equation solver in an indexed fashion. This solver, as the core algorithm of this paper, is reformulated in section \ref{sec4} as an operator given by a non-linear composition of matrices. Next, its role in transforming between pseudo-unitary groups and group-like objects is investigated.

\subsection{Dot product conservation formula in sets of Legendre-transformed functions}
    Legendre transforms are vastly used in Lagrangian Mechanics and Thermodynamics \cite{callen1998thermodynamics} to design a formalism that is analytically appropriate for the variables of a system. Consider $U=U(V_1, V_2, \cdots, V_n)$ a function of $n$ variables $V_i$ and $U[1] = U-p_1V_1$ its Legendre transform with respect to $V_i$.
    \begin{align} \label{lt1}
        U &= \sum_{i=1}^n p_i V_i
        \\\mathrm{d}U - p_1\mathrm{d}V_1
        &= \sum_{i=2}^n p_i\mathrm{d}V_i
        \\\mathrm{d}U - p_1\mathrm{d}V_1 - V_1\mathrm{d}p_1
        &= - V_1\mathrm{d}p_1 + \sum_{i=2}^n p_i\mathrm{d}V_i
        \\ \mathrm{d}(U-p_1V_1) &= - V_1\mathrm{d}p_1 + \sum_{i=2}^n p_i\mathrm{d}V_i
        \\ \mathrm{d} U[1] &= - V_1\mathrm{d}p_1 + \sum_{i=2}^n p_i\mathrm{d}V_i.
        \label{lt5}
    \end{align} 

        In matrix formalism, equations (\ref{lt1}) and (\ref{lt5}) can be expressed as dot products (equation (\ref{ldot})) between the vector $\mathbf{V}_U$ and its Legendre conjugate $\mathbf{V}^\star_U$ under the metric $g_U$.
        
        \begin{align} \label{ldot}
         U &= \mathbf{V}_U^\star g_U \mathbf{V}_U,
         \\ \text{with} \quad
            \mathbf{V}_{U} &= 
            \begin{bmatrix}
                V_1 \\ V_2 \\ \vdots \\ V_n
            \end{bmatrix}, \quad
            \mathbf{V}_{U}^\star =
            \begin{bmatrix}
                p_1 & p_2 & \cdots & p_n
            \end{bmatrix}.
        \end{align}

        Conservation of the dot product $U$ implies that $\mathrm{d}U$ in equation (\ref{consdot}) is zero.
        \begin{align} \label{consdot}
            \mathrm{d}U = \mathbf{V}_U^\star g_U \mathrm{d}\mathbf{V}_U
        \end{align}
	
    A Legendre transform over $U$ swaps entries between $\mathbf{V}_U$ and $\mathbf{V}^\star_U$, also changing the metric $g_U$ and resulting in a new dot product. The new metric (equation (\ref{rel3})) due to the Legendre transform over the $i$th variable of the vector $\mathbf{V}_U$ is obtained by extraction of the signs that accompany the products $p_iV_i$.
    \begin{align} \label{rel3}
        g_{U[n]}
        = (I-2|n\rangle \langle n|)g_U
    \end{align}

	The problem of a matrix representation of the transformation that maps a unitary operator to an $\eta$-pseudo-unitary one is the focus of this study. The reappearance of equation (\ref{rel3}), the dot product metric, is for now the main connection between the Legendre transform and the pseudo-unitary-metric transform to be presented in section \ref{kot}.

\subsection{Step-wise linear equation solver}
    Gauss-Jordan elimination is a traditional, robust, and still largely used matrix inversion algorithm due to its
    parallelization capability \cite{arfken}. It involves a parallel application of a sequence of row summations and
    row multiplications by constants in both a sample matrix and the identity matrix until the identity
    switches place with the sample-matrix. This way the matrix inverse is found.

    The Gauss-Jordan elimination algorithm is a type of linear equation solver for a system of the form of equation (\ref{lineqs}), where $\mathbf{x}$ is usually a vector of variables to be solved for and $\mathbf{v}$ is a vector of independent variables or constants. They are related by the coefficient matrix $Z$, whereas the solved equation (\ref{linsol}) relates $\mathbf{v}$ to $\mathbf{x}$ by the inverse of $Z$.
    \begin{align} \label{lineqs}
        \mathbf{v} &= Z \mathbf{x}
        \\ \mathbf{x} &= Z^{-1} \mathbf{v}
        \label{linsol}
    \end{align}

    The method for solving the system for only one entry of $\mathbf{x}$ involves isolation of a variable in a
row of the resulting system and its subsequent substitution in other rows. This algorithm $\hat{Y}_{ik}$ is summarized by a sequence
of four transformations throughout a matrix $Z$ (equations (\ref{e:pialgo}) through (\ref{e:pialgo2})), with $i, k$ fixed and $\sigma, \rho$ changing. The $i$th entry of the output-vector is swapped with the $k$th entry of the input-vector in the process.

        \begin{align}
        \label{e:pialgo}
            Z_{ik}         &\to Z_{ik}^{-1} ;
            \\ Z_{i\sigma} &\to -Z_{ik}^{-1} Z_{i\sigma},  \quad \text{for all} \quad \sigma \neq k;
            \\ Z_{\rho k} &\to Z_{\rho k}Z_{ik}^{-1}, \quad \text{for all} \quad \rho \neq i;
            \\ Z_{\rho \sigma} &\to Z_{\rho \sigma} -  Z_{\rho k} Z_{ik}^{-1} Z_{i\sigma}, \quad \text{for all} \quad \sigma \neq k,  \rho \neq i.
            \label{e:pialgo2}
        \end{align}

    This well-known procedure is the core algorithm to transit between a family of pseudo-unitary groups. The primarily numerical approach of equations (\ref{e:pialgo}) through (\ref{e:pialgo2}) was reformulated in a semi-analytical fashion for clarity, followed by a deeper study of its representations, potential applications and experimental implications.
    
\section{\label{sec4}Results and Discussion}

    \subsection{Correspondence between partial inversion, permutation and total matrix inversion}

We use $\hat{Y}_{ik}$ to represent an application of  equations (\ref{e:pialgo}) through (\ref{e:pialgo2}), and adopt the name "partial inversion" to refer to it throughout this paper, since it can be used to invert matrices. Its matrix representation is obtained from a sum of deformed detachments, as illustrated by the matrix grids in  equation (\ref{detachments}) that represent the adding terms of  equation (\ref{defdeta0}) for the case of $\hat{Y}_{22}$ over a square matrix of dimension $4$. The grids take  $\cdot$ for unchanged elements, $\bullet$ for non-zero changed elements and empty entries for zeros.

\setlength{\tabcolsep}{1pt}
\begin{align} \label{detachments}
  \begin{tabular}{l} \detY \end{tabular}
    =
  \begin{tabular}{l}  \detI \end{tabular}
  +
  \begin{tabular}{l}  \detZC \end{tabular}
  -
  \begin{tabular}{l}  \detCZ \end{tabular}
  +
  \begin{tabular}{l}  \detCZC \end{tabular}
  -
  \begin{tabular}{l}  \detZCZ \end{tabular}.
\end{align}

\begin{align} \label{defdeta0}
    \hat{Y}_{ik} Z &= Z + ZC_{kk} - C_{ii}Z + C_{ii}ZC_{kk} - ZC_{ik}^\mathrm{T}Z,
    \\ \notag &\text{with}
    \quad C_{ik}= \frac{|i \rangle \langle k |}{\langle i|Z| k\rangle}.
\end{align}

If $i=k$,  equation (\ref{defdeta0}) becomes  (\ref{defdeta}), which reduces to the operator defined in equation (\ref{pia}), with the appearance of equation (\ref{ybe}), notoriously similar to the Yang-Baxter equation from Braid Theory \cite{kauf}, except for the transposition.

    \begin{align} \label{defdeta}
        \hat{Y}_{ii} Z =  Z + ZC_{ii} - C_{ii}Z + C_{ii}ZC_{ii} - ZC_{ii}^\mathrm{T}Z, 
        \\ \notag \text{with} \quad 
        C_{ii} = \frac{|i \rangle \langle i| }{\langle i|Z |i \rangle }.
    \end{align}

        \begin{align} \label{pia}
            \hat{Y}_{ii} &=  \hat{I} + \widehat{[(\quad) ,C_{ii}]} 
            + \widehat{[(\quad),C_{ii}]}_{\star}, 
            \\ \notag \text{with} \quad C_{ii} &= \frac{|i\rangle \langle i| }{\langle i|(\quad)|i \rangle },
            \\ [A,B] &= AB - BA,
            \\ [A,B]_\star &= BAB - AB^\mathrm{T}A. \label{ybe}
        \end{align}

Automatically from the permutation of input and output vectors, any sequence of $\hat{Y}_{ii}$ for all $i$ in a matrix  leads to its total inverse. The occurrence of permutations is exemplified in equation (\ref{here}) for 2-vectors, where $\pi_1$ is a permutation matrix.
\begin{align} \label{here}
\begin{pmatrix} x \\ y \end{pmatrix}
&= Z \begin{pmatrix} a \\ b \end{pmatrix}
, \\ \notag
\begin{pmatrix} x \\ y \end{pmatrix}
&=(Z\pi_1) \begin{pmatrix} b \\ a \end{pmatrix}
, \\ \notag
\begin{pmatrix} a \\ y \end{pmatrix}
&=  \hat{Y}_{00}(Z)  \begin{pmatrix} x \\ b \end{pmatrix}.
\end{align}

The partial inversion group $G(\hat{Y}_{ik},n)$ of $n$-dimensional square matrices contains and is larger than the general linear group $GL(n)$, since $G(\hat{Y}_{ik})$ contains the operation of total matrix inverse. The group $G(\hat{Y}_{ik})$ also contains representations of the symmetric group $S(n)$ and the Unitary group $U(n)$ (section \ref{secunitrep}). The relations that define generators of the symmetric group are defined in equations (\ref{third0}) through (\ref{third})), with a chained cycle of indices, such as $\hat{Y}_{i,i+1}$, $\hat{Y}_{i+2,i+1}$, $\hat{Y}_{i+2,i+3}$, $\cdots$, as generators. The commutation in equation (\ref{third}) expresses that permutations that are not chained are independent of order.
    
    \begin{align}
            \label{third0}
        \hat{Y}_{ik}\hat{Y}_{ik} &= \hat{I},
        \\\label{third1} \hat{Y}_{i-1,i} \hat{Y}_{i+1,i} \hat{Y}_{i-1,i} &= 
        \hat{Y}_{i+1,i} \hat{Y}_{i-1,i} \hat{Y}_{i+1,i},
        \\ [\hat{Y}_{ik}, \hat{Y}_{j\ell}] &= 0 \, \iff \, |i-j| >0, \, |k-\ell| >0.
        \label{third}
    \end{align}
    
        The correct partial inversion that maps an S-matrix to a T-matrix depends on the conventions used for both S-matrix and T-matrix. From the first presentation of $S$ in equation (\ref{sss}), other possible conventions use $r$ in place of $t$, or $r^\prime$ instead of $-r^\prime$ (cf. \cite{snyman2008calculation}). 
        
        In terms of $\hat{Y}_{ik}$ over the transfer matrix $T$, the $SU(2)$ S-matrix is read as equation (\ref{jok}).
        \begin{align} \label{jok}
            S &= \hat{Y}_{11}T
            \\ \notag  &=  T + TC_{11} - C_{11}T + C_{11}TC_{11} - TC_{11}^\mathrm{T} T
            \\ \notag &=
            \left[\begin{matrix}\frac{1}{t^*} & \frac{r}{t^*}\\\frac{r^*}{t^*} & \frac{1}{t^*}\end{matrix}\right]
                    +
            \left[\begin{matrix}0 & r\\0 & 1\end{matrix}\right]
                    -
            \left[\begin{matrix}0 & 0\\r^* & 1\end{matrix}\right]
                    +
            \left[\begin{matrix}0 & 0\\0 & t^*\end{matrix}\right]
                    -
            \left[\begin{matrix}\frac{r r^*}{t^*} & \frac{r}{t^*}\\\frac{r^*}{t^*} & \frac{1}{t^*}\end{matrix}\right]
            \\ \notag &=
            \begin{pmatrix} 
              t & r
           \\ -r^* & t^*
           \end{pmatrix}.
        \end{align}

         Equations (\ref{eqst}) and (\ref{eqst2})  exemplify other applications of partial inversion on the $SU(2)$ S-matrix. For higher dimensions, since the S and T-matrices are composed by four blocks of matrices, equations (\ref{e:pialgo}) through (\ref{e:pialgo2}) should be used with the appropriate order of matrix multiplication, using the blocks in place of the entry $Z$.
        \begin{align} \label{eqst}
        A &= \hat{Y}_{01} \hat{Y}_{00}S =
		\frac{1}{r}        
        \begin{pmatrix}
        -1 & t^* \\ -t & 1
        \end{pmatrix},
        \\ \label{eqst2}
        B &= \hat{Y}_{01} S = \frac{1}{r}  
        \begin{pmatrix}
        -t & 1 \\ -1 & t^*
        \end{pmatrix}.
        \end{align}
        
    An example closer to experimental realization is illustrated by the matrix representation of the half-wave plate, a gadget that rotates the polarization of the electromagnetic oscillation planes of a photon by an angle $2\theta$. This gadget, represented by equation (\ref{pika}), reproduces the Hadamard gate in polarization space at $\theta=\pi/8$ \cite{hecht}.  Equation (\ref{pika2}) shows an example of a $\sigma_3$-pseudo-unitary version of it.
    \begin{align} \label{pika}
    C(\theta) &=
    \begin{bmatrix}
    \cos(2\theta)     &\sin(2\theta)
    \\  \sin(2\theta) &-\cos(2\theta)
    \end{bmatrix},
    \\ \label{pika2} C^\prime (\theta) = \hat{Y}_{11} C(\theta)
    &=
    \begin{bmatrix}
    \sec(2\theta) & -\tan(2\theta)
   \\ \tan(2\theta)  & -\sec(2\theta)
    \end{bmatrix}.
    \end{align}
    
    Similarly to how the $X,Y,Z$ gates are generators of the the Pauli group as both unitary and Hermitian matrices, $W$ in equation (\ref{zzpauli}) is the general solution to the constraints for a $\sigma_3$-pseudo-Hermitian and $\sigma_3$-pseudo-unitary matrix. 
    
 \begin{align}\label{zzpauli}
    W(z,\theta) = \begin{pmatrix}
        \sqrt{1+z^2}       & z\mathrm{e}^{\mathrm{i}\theta}
        \\ -z\mathrm{e}^{\mathrm{i}\theta} & -\sqrt{1+z^2}
    \end{pmatrix}, \quad \mathrm{with} \quad z,\,\theta \in \mathbb{R}.
 \end{align}

        
    
    \subsection{Diagrammatic representation}
    		Since permutations have a plane diagrammatic representation \cite{kauf}, a natural extension to partial inversions as permutations through graph time requires an extra dimension. In the proposed representation, we omit the straight (unaltered) strands common to permutation and braid diagrams to ease the tridimensional visualization. The bottom tile is the initial matrix, with permutations growing upwards.

\begin{align}
 \begin{tabular}{c}
\includegraphics[]{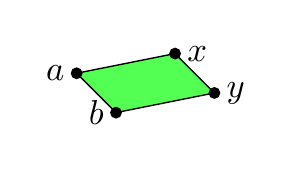}
\end{tabular}
& \leftrightarrow &
    \begin{pmatrix}x \\ y \end{pmatrix} = M
    \begin{pmatrix}a \\ b\end{pmatrix}.
\end{align}

\begin{align}
\begin{tabular}{c}
\includegraphics[]{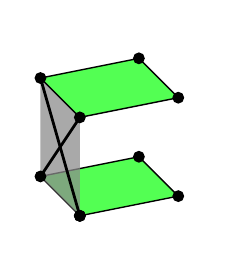}
\end{tabular}	    
    = M\hat{\pi}_{1},
\end{align}

\begin{align}
\begin{tabular}{c}
\includegraphics[]{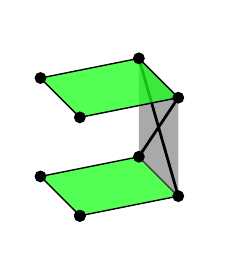}
\end{tabular}
 = \hat{\pi}_{1}M, 
\end{align}

\begin{align}
\begin{tabular}{c}
\includegraphics[]{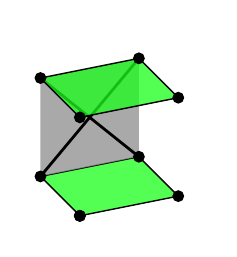}
\end{tabular}
= \hat{Y}_{00}M,
\end{align}

\begin{align}
\begin{tabular}{c}
\includegraphics[]{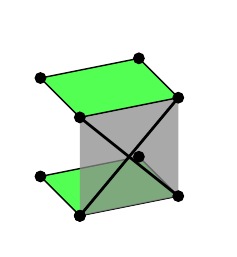}
\end{tabular}	    
&= \hat{Y}_{11}M,   
\\ \begin{tabular}{c}
\includegraphics[]{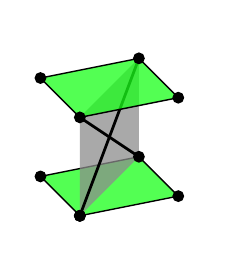}
\end{tabular}
&= \hat{Y}_{01}M,
\\ \begin{tabular}{c}
\includegraphics[]{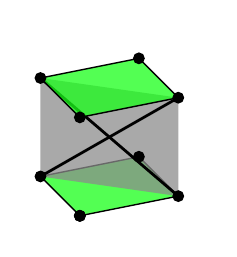}
\end{tabular}
&= \hat{Y}_{10}M.
\end{align}

Each pair of parallel or crossing planes is a commutative pair, therefore they can be applied simultaneously.

\begin{align} \notag
\begin{tabular}{c}
\includegraphics[]{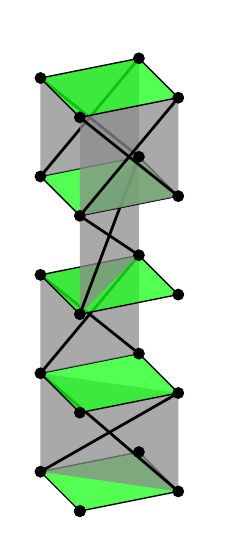}
\end{tabular}	    
=
\begin{tabular}{c}
\includegraphics[]{tow12}
\end{tabular}	    
\\ \leftrightarrow \quad \hat{Y}_{11}\hat{Y}_{00}\hat{Y}_{01}\hat{Y}_{00}\hat{Y}_{10} M
=M \hat{\pi}_{1},
\end{align}

\begin{align} \notag
\begin{tabular}{c}
\includegraphics[]{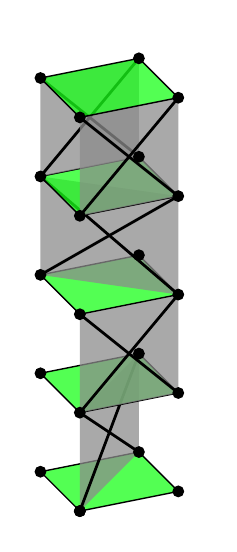}
\end{tabular}	    
=   
\begin{tabular}{c}
\includegraphics[]{tow32}
\end{tabular}	    
\\ \leftrightarrow \quad \hat{Y}_{11}\hat{Y}_{00}\hat{Y}_{10}\hat{Y}_{11}\hat{Y}_{01} M
=\hat{\pi}_{1} M,
\end{align}

\begin{align} \notag
\begin{tabular}{c}
\includegraphics[]{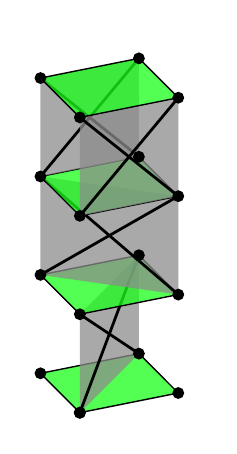}
\end{tabular}	    
=
\begin{tabular}{c}
\includegraphics[]{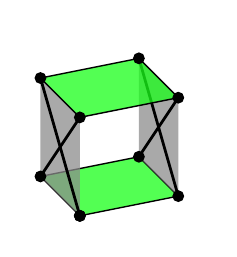}
\end{tabular}
\\ 
\leftrightarrow \quad \hat{Y}_{11}\hat{Y}_{00}\hat{Y}_{10}\hat{Y}_{01} M
= \hat{\pi}_{1}M\hat{\pi}_{1},
\end{align}

\begin{align} \notag &
\begin{tabular}{c}
\includegraphics[]{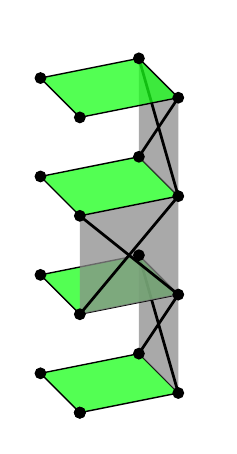}
\end{tabular}	    
=
\begin{tabular}{c}
\includegraphics[]{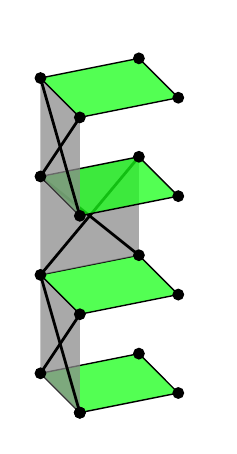}
\end{tabular}
=
\begin{tabular}{c}
\includegraphics[]{tow53}
\end{tabular}
\\ &\leftrightarrow \quad 
\hat{\pi}_{1} \hat{Y}_{11} (\hat{\pi}_{1}M)
=
[\hat{Y}_{00} (M\hat{\pi}_{1})]\hat{\pi}_{1}
= \hat{Y}_{01}M,
\end{align}

\begin{align} \notag &
\begin{tabular}{c}
\includegraphics[]{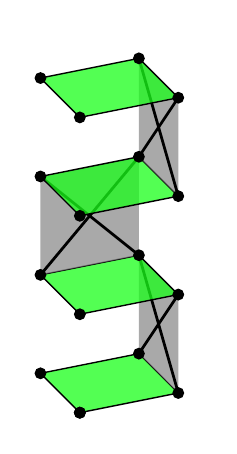}
\end{tabular}	    
=   
\begin{tabular}{c}
\includegraphics[]{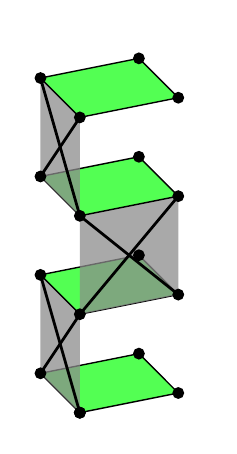}
\end{tabular}
=
\begin{tabular}{c}
\includegraphics[]{tow43}
\end{tabular}
\\
&\leftrightarrow \quad 
\hat{\pi}_{1} \hat{Y}_{00} (\hat{\pi}_{1}M)
=
[\hat{Y}_{11} (M\hat{\pi}_{1})]\hat{\pi}_{1}
= \hat{Y}_{10}M.
\end{align}

    \subsection{Deformed algebra of pseudo-unitary metric operator} \label{kot}
        As exemplified in the S-matrix case, the application of partial inversion to a matrix entry of a matrix representation of an $\eta_0$-pseudo-unitary group maps it to a different, $\eta_1$-pseudo-unitary group. This is a direct consequence of the non-commutativity of some partial inversions with a similarity transformation $\hat{\eta}$.  Equation (\ref{notcom}) fully describes the problem of defining $\eta_1$ in terms of $\eta_0$ for an $\eta_0$-pseudo-unitary matrix $Z$. 
        In contrast, the sequence of partial-inversions that amounts to total inversion $\hat{y}_1$ commutes with similarity transformations (equation (\ref{commutes})). Here $\hat{\kappa}_1 Z = Z^\dagger$.
        
        \begin{align}
		\label{notcom}
		\hat{Y}_{ik} \hat{y}_1 \hat{\kappa}_1 \hat{\eta}_0 U
		&= \hat{y}_1 \hat{\kappa}_1 \hat{\eta}_1 \hat{Y}_{ik} U
		\\ \notag \leftrightarrow
		\quad
		\hat{Y}_{ik} Z &= 
		\hat{\eta_1 }^{-1} \hat{\kappa}_1 \hat{y}_1 \hat{Y}_{ik}
		\hat{y}_1 \hat{\kappa}_1 \hat{\eta}_0 Z.
		\end{align}
        
        \begin{align}
		\label{commutes}
        \hat{y}_1 \hat{\eta} Z &= \hat{\eta} \hat{y}_1 Z  
        \\ \notag \leftrightarrow \quad  (\eta^{-1} Z \eta )^{-1} &= \eta^{-1} Z^{-1} \eta.
        \end{align}

        The non-commutativity relation can be recast as a deformed commutator, taking $\hat{\eta}Z=\eta^{-1} Z \eta$ as a metric operator acting over an arbitrary matrix $Z$.
        Consider the deformed commutator in equation (\ref{deformedcom}), with $\hat{\eta}_1 = \hat{\Omega} \hat{\eta}_0$ the new pseudo-unitary metric operator obtained after a partial inversion. $\hat{\Omega}$ is the deformation metric operator.
        \begin{align}
        \label{deformedcom}
            [\hat{Y}_{ik}, \hat{\eta}_0 ]_\Omega 
            = \hat{Y}_{ik} \hat{\eta}_0 - \hat{\Omega} \hat{\eta}_0 \hat{Y}_{ik}
            =0.
        \end{align}

        Opening the terms of  equation (\ref{deformedcom}) to seek the general solution for $\hat{\eta}_1$ in terms of $\eta_0$ and $(i,k)$ results in a non-analytical and hardly tractable equation. Fortunately, numerical evaluations based on representations of the dihedral group are easily obtained following the occurrence of $\hat{\eta}_1=\hat{\sigma}_3$ for the T-matrix from an $SU(2)$ S-matrix.         

        We use as representations of the dihedral group $\hat{\tau}_i$ for transposition around the $i$th axis (also diagonal), $\hat{\kappa}_i$ for complex conjugation with transposition and $\hat{y}_i$ for inversion along the $i$th diagonal. They obey equations (\ref{rels}) through (\ref{rels3}) (in place of $\hat{a}$).
        \begin{align}
        \label{rels}
            \hat{a}_{i} \hat{a}_{i+1} &= \hat{a}_{i+2},
            \\ \hat{a}_i \hat{a}_i &= \hat{I},
            \\ [\hat{a}_i, \hat{a}_j] &=0.
            \label{rels3}
        \end{align}

Although Pauli matrices in equation (\ref{paulig}) anti-commute (equation (\ref{paulig2})), their similarity transformations $\hat{\sigma}_i(\quad) = \sigma_i(\quad )\sigma_i$ do commute and form a dihedral group (equation (\ref{paulig3})).
        \begin{align}
            \label{paulig}
            \sigma_1 = \begin{pmatrix} 0 & 1 \\ 1 & 0 \end{pmatrix}
            ,\quad \sigma_2 = \begin{pmatrix} 0 & -i \\ i & 0 \end{pmatrix}
            ,\quad \sigma_3 = \begin{pmatrix} 1 & 0 \\ 0 & -1 \end{pmatrix}.
            \end{align}
            
            \begin{align}
             [\sigma_i, \sigma_{i+1}] &= 2 \mathrm{i} \sigma_{i+2} 
            \label{paulig2},
            \\ \sigma_i \sigma_{i+1} Z \sigma_{i+1} \sigma_{i} &= 
            \mathrm{i}\sigma_{i+2} Z (-\mathrm{i})\sigma_{i+2} =
            \sigma_{i+2} Z \sigma_{i+2}
            \label{paulig3}.
        \end{align}

        Transposition along the main diagonal is denoted by an operator whose matrix entries are orthonormal functions $F_{\mu\nu}(\theta)$ under integration, therefore $\int \mathrm{d}\theta F_{\mu\nu}F_{\lambda \rho} = \delta_{\mu\lambda} \delta_{\nu\rho}$, serving the purpose of permuting entries around the main diagonal. 
        The three representations of $D_2(n)$ (equations (\ref{reps}) through (\ref{reps2}),  equation (\ref{reps3}), and $\hat{\kappa}_i$ as for $\hat{\tau}_i$ with a complex conjugation operator) have the common element $\hat{\tau}_3$, i.e., $D_2(\hat{\tau}) \cap D_2(\hat{y}) \cap D_2(\hat{\sigma}) \cap D_2(\hat{\kappa}) = \hat{\tau_3}$. 
        \begin{align} \label{reps}
             \hat{\tau}_1 &= \int \mathrm{d} \theta \begin{pmatrix} F_{00} & F_{01} \\ F_{10} & F_{11} \end{pmatrix} \left(\, \right)
            \begin{pmatrix} F_{00} & F_{01} \\ F_{10} & F_{11} \end{pmatrix}
            ,\\ \hat{\tau}_2 &= \hat{\tau_3}\hat{\tau_1}
            ,\\ \hat{\tau}_3 &= \begin{pmatrix} 0  &1 \\ 1 & 0 \end{pmatrix}
            (\, )
            \begin{pmatrix} 0  &1 \\ 1 & 0 \end{pmatrix},
        \label{reps2}   \end{align}
            
            \begin{align}
             \hat{y}_1 = \prod_{i=0}^{N-1} \hat{Y}_{ii}
            ,\quad \hat{y}_2 = \prod_{i=0}^{N-1} \hat{Y}_{i,N-i-1},
            \quad  \hat{\tau_3} = \hat{y}_1 \hat{y}_2.
            \label{reps3}
        \end{align}

As consequence, any product of them can exchange $\hat{\tau}_3$ to change axes (e.g., equation (\ref{hoa})).
\begin{align} \label{hoa}
\hat{\sigma}_2 \hat{\tau}_1 = \hat{\sigma}_2 \hat{\tau}_3 \hat{\tau}_2 =
\hat{\sigma}_3 \hat{\tau}_2, \quad \hat{\sigma}_1 =\hat{\tau}_3.
\end{align}

       Running a routine in python to check if any combination of the above-mentioned metric operators is a candidate for the metric operator of a pseudo-unitary group, taking the $SU(2)$ S-matrix as starting point, one finds that this example of S-matrix itself pertains to three special pseudo-unitary groups, and so do other groups and group-like structures derived from it by partial inversion. 
       It is, therefore, convenient to use  equation (\ref{cuppy}) to represent the group of a metric vector $\vec{\alpha}$ as the intersection of the groups of each vector entry. In our case, the metric vectors are 3-vectors, and only two-dimensional matrices are studied.
       \begin{align} \label{cuppy}
       SU(2, \vec{ \hat{\alpha } } ) 
       = 
       SU(2, \hat{ \alpha }^{(0)} ) 
       \cap 
       SU(2, \hat{\alpha }^{(1)} ) 
       \cap
       SU(2, \hat{\alpha }^{(2)} ).
       \end{align}

\begin{figure}[ht]  \centering
	\caption{
        Graph that defines the relation between $U(2^n,I^{\otimes n})$ and $U(2^n,\sigma_3^{\otimes n})$ (magenta link between nodes of respective metrics), and defines $SU(2,\eta)$ metric-deformation operators in $\hat{\Omega}(i,k,\nu,\rho)$ (full black and red edges) as links between metric vectors (nodes $\vec{ \hat{ \alpha}} $, $\vec{ \hat{\phi} }$). $\vec{ \hat{\phi} }$ and $\vec{ \hat{\epsilon} }$ are linked by powers, not $\hat{\Omega}$ (dashed edges). The magenta edge marks a sequence of equal indices as the first $2^{n-1}$ odious numbers in the odious series. Black edges mark $i=k$, and red edges mark $i\neq k$, with $i,k$ the partial inversion indices. The nodes of pseudo-unitary metrics are linked to the nodes of pseudo-special-unitary metrics by dotted edges to represent their relation with simpler and more symmetric sets of metrics.\label{hex}}
	\includegraphics[width=0.5\textwidth]{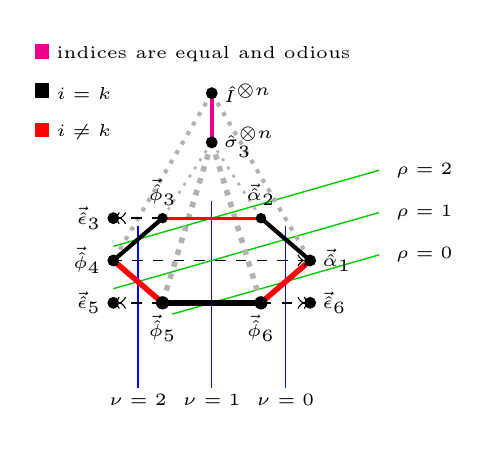}
\end{figure}

        In figure \ref{hex}, the upper region of the graph with two nodes linked by a magenta edge marks a map between pseudo-unitary groups of higher dimensions that are grown by tensor product from elements of  $U(2, \hat{I})$ and $U(2, \hat{\sigma}_3)$, respectively. These groups were numerically found to be mapped to each other by a sequence of partial inversions with the first $2^{n-1}$ odious numbers as indices, where $2^n$ is the dimension of the group element.
        
        Odious numbers are integers whose binary representations have odd numbers of "1"s. They are object of study of Number Theory and are of rare occurrence in mathematical problems, along with their reciprocal set of evil numbers, the integers with even number of "1"s \cite{allouche2016beyond}. 
        The first ten numbers of the odious sequence $\mathbb{O}$ are given in equation (\ref{odse}).
        \begin{align} \label{odse}
            \mathbb{O} = \{1, 2, 4, 7, 8, 11, 13, 14, 16, 19, \cdots\}
        \end{align}

        The first three examples of the map between the two pseudo-unitary groups in figure \ref{hex} are given in equations (\ref{exz}) through (\ref{exzz}). The matrix indices are conveyed to start from $0$.
        \begin{align} \label{exz}
            U(2,\hat{\sigma}_3) &= \hat{Y}_{11} U(2,\hat{I}_2)
            \\ 
            U(4,\hat{\sigma}_3^{\otimes 2}) &= \hat{Y}_{22}\hat{Y}_{11} U(4,\hat{I}_2^{\otimes 2})
            \\ 
            U(8,\hat{\sigma}_3^{\otimes 3}) &= \hat{Y}_{77}\hat{Y}_{44}\hat{Y}_{22}\hat{Y}_{11} U(8,\hat{I}_2^{\otimes 3})
            \label{exzz}
        \end{align}

		Partial inversions do not always produce new groups. In figure \ref{hex}, $\vec{ \hat{\alpha } }_i$ are metric vectors of pseudo-unitary groups, $\vec{ \hat{\phi } }_i$ are group-like structures that obey the pseudo-unitary inversion constraint, but are made only of odd powers. Their even powers do form pseudo-unitary groups and are marked on the diagram by $ \vec{ \hat{\epsilon } }_i$ and $ \vec{ \hat{\alpha } }_1$ nodes.
		
\begin{align} \label{large1}
 \vec{ \hat{\alpha } }_1
 &= \begin{pmatrix} \hat{I} \\ \hat{\sigma}_3 \hat{\tau}_1 \hat{\kappa}_2 \\ \hat{y}_1 \hat{\tau}_2 \hat{\sigma}_3 \end{pmatrix}, 
 \quad 
  \vec{ \hat{\alpha } }_2
 = \begin{pmatrix} \hat{\sigma}_3 \\ \hat{\sigma}_3 \hat{\tau}_2 \\ \hat{y}_1 \hat{\tau}_1 \hat{\kappa}_2 \end{pmatrix}, 
 \\ 
   \vec{ \hat{\phi } }_3
 &= \begin{pmatrix} -\hat{\tau}_2 \\ \hat{\sigma}_3 \\ -\hat{y}_1 \hat{\tau}_1 \hat{\kappa}_2 \hat{\sigma}_3 \end{pmatrix}, 
 \quad 
    \vec{ \hat{\phi } }_4
= \begin{pmatrix}
- \hat{\sigma_3} \hat{\tau}_1 \hat{\kappa}_2 \\ \hat{I} \\ -\hat{y}_1 \hat{\sigma}_3 \hat{\tau}_2
\end{pmatrix} ,
\\ \label{large2}    \vec{ \hat{\phi } }_5
&= \begin{pmatrix}
- \hat{\sigma_3} \hat{\tau}_1 \\ -\hat{\sigma}_3 \\ \hat{y}_1 \hat{\tau}_1 \hat{\kappa}_1
\end{pmatrix}, \quad     
    \vec{ \hat{\phi } }_6
= \begin{pmatrix}
- \hat{\sigma_3} \\ \hat{\tau}_1 \\ -\hat{y}_1 \hat{\sigma}_2 \hat{\tau}_1 \hat{\kappa}_2
\end{pmatrix}, 
\\
    \vec{ \hat{\epsilon } }_5
&= \begin{pmatrix}
 \hat{\sigma_3} \\ \hat{\tau}_1 \hat{\sigma}_3 \\ \hat{y}_1 \hat{\tau}_1 \hat{\kappa}_1
\end{pmatrix}, \quad
    \vec{ \hat{\epsilon } }_6
= \begin{pmatrix}
 \hat{\sigma_3} \\ \hat{\tau}_1 \\ \hat{y}_1 \hat{\sigma}_2 \hat{\tau}_1 \hat{\kappa}_2
\end{pmatrix}.
\label{largelast}
\end{align}		
		In addition to  equations (\ref{large1}) through (\ref{largelast}), $\vec{\hat{\epsilon}}_3$ is highly constrained and contains all possible combinations of $D_2$ generators as vector entries. It is worth noting that all entries containing $\hat{y}_1$ express a symmetry property (equation (\ref{sp})). For example, if $\hat{\eta}=\hat{a}\hat{y}_1$, then 
		\begin{align} \label{sp}
		\hat{y}_1 Z &= \hat{\eta} \hat{\kappa}_1 Z = \hat{y}_1 \hat{a} \hat{\kappa}_1 Z
		\quad \iff \quad Z = \hat{a} \hat{\kappa}_1 Z.
		\end{align}

		This heuristic graph in figure \ref{hex} allows us to define the deformation matrix $\hat{\Omega}$ that maps one metric vector to another, instead of trying to solve  equation (\ref{deformedcom}). It depends both on the partial inversion indices $i,k$ and on the graph coordinates $\nu,\rho$. $\nu$ can be interpreted as the distance from the unitary group (defined within $\vec{\hat{\alpha}} _1$), whereas $\rho$ is a rolling number, and the superscript notation $[\rho]$ is used to roll matrix entries along the main diagonal.

\begin{align}
\hat{\Omega}(i,k,\nu,\rho) = \begin{cases}
P^{[\rho]}(\hat{\sigma}_3, \hat{\kappa}^{\nu} \hat{\kappa}_1 ), \quad  \text{if} \quad i=k;
\\ 
P^{[\rho]}(
\hat{\kappa}^{\nu} \hat{\kappa}_2,
-\hat{\sigma}_3
 ), \quad  \text{otherwise.}
\end{cases}
\end{align}
with $\hat{\kappa} Z = Z^*$ the complex conjugation operator over the matrix $Z$, $\hat{\kappa}^2= \hat{I}$, and
\begin{align} \label{pmatrices}
P^{[0]}(\hat{a},\hat{b})&= \begin{pmatrix}
\hat{b} &0&0 \\ 0& \hat{a}\hat{b} &0 \\ 0& 0&\hat{a} 
\end{pmatrix},
\\ \notag P^{[1]}(\hat{a},\hat{b})&= \begin{pmatrix}
\hat{a} &0&0 \\ 0& \hat{b} &0 \\ 0& 0&\hat{a}\hat{b} 
\end{pmatrix},
\\ \notag P^{[2]}(\hat{a},\hat{b})
&= \begin{pmatrix}
\hat{a}\hat{b} &0&0 \\ 0& \hat{a} &0 \\ 0&0 &\hat{b}
\end{pmatrix}.
\end{align}

As a first example, equations (\ref{ex1}) show the effect of partial inversion on the metric vector from the S-matrix group $SU(2, \vec{\hat{\alpha}}_1 )$ to the T-matrix group $SU(2, \vec{\hat{\alpha}}_2 )$.

\begin{align} \label{ex1}
\hat{Y}_{11} S = T \quad \iff \quad
\hat{\Omega}(1,1,0,1) 
\vec{ \hat{\alpha} }_1 &= \vec{ \hat{\alpha} }_2,
\\ \notag
P^{[1]}(\hat{\sigma}_3, \hat{\kappa}_1 ) 
\vec{ \hat{\alpha} }_1 &= \vec{ \hat{\alpha} }_2,
\\ \notag
\begin{pmatrix}
   \hat{\sigma}_3 &&
\\ & \hat{\kappa}_1 &
\\ && \hat{\kappa}_1 \hat{\sigma}_3
\end{pmatrix}
\begin{pmatrix}
\hat{I}
\\ \hat{\sigma}_3 \hat{\tau}_1 \hat{\kappa}_2
\\ \hat{y}_1 \hat{\tau_2} \hat{\sigma}_3
\end{pmatrix}
&= \begin{pmatrix}
\hat{\sigma}_3
\\  \hat{\sigma}_3 \hat{\tau}_2
\\ \hat{y}_1 \hat{\tau}_1 \hat{\kappa}_2
\end{pmatrix}.
\end{align}

\subsection{Unitary representation of a sequence of partial inversions\label{secunitrep}}
    A powerful consequence of figure \ref{hex} is that, for $Z\in SU(2, \hat{I}_2)$ there special sequences of partial inversions (around the hexagon graph in figure \ref{hex}) with $SU(2)$ representations, and similarly for the $\sigma_3$-pseudo-special unitary group, as shown in equation (\ref{nont}).
\begin{align} \label{nont}
Z \in &SU(2) 
 \iff
\hat{Y}_{f,f^\prime} \hat{Y}_{ee} \hat{Y}_{d,d^\prime} \hat{Y}_{cc} \hat{Y}_{b,b^\prime} \hat{Y}_{aa} Z \in SU(2),
\\ \notag &\text{with} \quad a,b, b^\prime, c,d, d^\prime,e,f, f^\prime \in \{0,1\},
\\ \notag &\text{and with} \quad  b^\prime \neq b, \, d^\prime \neq d, \, f^\prime \neq f.
\end{align}

Equation (\ref{nont}) means that it is possible to define a $SU(2)$ matrix representation for sequences of partial inversions that obey that constraint. The example in equations (\ref{exuni}) through (\ref{exuni3}) emphasizes that such representation depends on the entries of the target matrix $Z$, but a universal representation can still be obtained as long as $Z$ is a general, symbolic $SU(2)$ matrix.

\begin{align} \label{exuni}
    Z^\prime = \hat{Y}_{01}\hat{Y}_{00}\hat{Y}_{10}
    \hat{Y}_{00}\hat{Y}_{01}\hat{Y}_{11}
    Z
\end{align}
\begin{align} \label{exuni2}
    \hat{Y}_{01}\hat{Y}_{00}\hat{Y}_{10}
    \hat{Y}_{00}\hat{Y}_{01}\hat{Y}_{11} &= Z^\prime Z^{-1}
    \\\notag &= \begin{pmatrix}
           (t^{*})^2+(r^{*})^2 & -(rt^* + r^*t)
        \\  rt^* + r^*t & t^2+r^2
    \end{pmatrix},
\end{align}

\begin{align} \label{exuni3}
    \text{with} \quad Z = \begin{pmatrix}
        t & r
        \\-r^* & t^*
    \end{pmatrix}.
\end{align}

    In this example, the unitary representation is simply obtained by implementing the transformations $t\to (t^{*})^2+(r^{*})^2$ and $r \to -(rt^* + r^*t)$.

    \subsection{Dot-product-conservation metric}
        The pseudo-unitary metric ($\eta$) also defines a pseudo-unitary dot-product \cite{znojil}. One can define a conservative dot-product (from a previously conservative one) straightforwardly from the change of sign due to swaps of vector entries between input and output vector in the partial inversion (equation (\ref{dpc})).
        
        \begin{align}
            \label{dpc}
            \psi_1^* \psi_1 + \psi_2^* \psi_2 &= \psi_3^*\psi_3 + \psi_4^*\psi_4 \quad  
            \\ \notag \leftrightarrow 
            \quad \psi_1^*\psi_1 - \psi_3^*\psi_3 &= -\psi_2^*\psi_2 + \psi_4^*\psi_4.
        \end{align}
        
        The final input ($g_\text{in}^\prime$) and output ($g_\text{out}^\prime$) dot-product conservation metrics depend explicitly on the partial inversion indices $i,k$ and the initial input and output metrics $g_\text{in}, g_\text{out}$ through equation (\ref{ifg}).
        
        \begin{align} \label{ifg}
            &\text{If} \, g_\alpha(i,i)=g_\alpha(k,k):
            \\ \notag
            &\quad \quad   g_\text{in}^\prime (i,k) = 
            \left(I
            - 2 |k \rangle \langle k| \right) g_\text{in}, 
            \\ \notag
            &\quad \quad   g_\text{out}^\prime (i,k) = 
            \left(I
            - 2 |i \rangle \langle i| \right) g_\text{out}, 
            \\ \notag &\text{else:}
            \\ \notag &\quad \quad  g_\alpha^\prime =g_\alpha.
        \end{align}

    Let us start an example from the S-matrix, whose vector space obeys the dot-product conservation law under the identity metric $I$, i.e. $g_\text{in} = g_\text{out} = I $, (equations (\ref{slaw0}) through (\ref{lasteq}), shortened from the previously defined $\hat{I} \in \vec{\hat{\alpha} }_1$).
    \begin{align} \label{slaw0}
        S &\in U(n,\hat{\eta}=\hat{I}).
        \\ \begin{bmatrix} \psi_{01} \\ \psi_{10} \end{bmatrix}
        &= S \begin{bmatrix} \psi_{00} \\ \psi_{11} \end{bmatrix}.
        \\ 
        \begin{bmatrix} \psi_{01}^* & \psi_{10}^* \end{bmatrix}
        \begin{pmatrix}
           1 & 0
        \\ 0 & 1
        \end{pmatrix}
        \begin{bmatrix} \psi_{01} \\ \psi_{10} \end{bmatrix}
        &=     
        \begin{bmatrix} \psi_{00}^* & \psi_{11}^* \end{bmatrix}
        \begin{pmatrix} 1 & 0 \\ 0 & 1 \end{pmatrix}
        \begin{bmatrix} \psi_{00} \\ \psi_{11} \end{bmatrix}.
        \label{lasteq}
    \end{align}
    
    An application of $\hat{Y}_{11}$ introduces the new pseudo-unitary metric coincidentally as the new dot-product-conservation metric ($\eta= g_\text{in} = g_\text{out} = \sigma_3$, shortened from $\sigma_3 \in \vec{\hat{\alpha}}_2$). Since equations (\ref{ifg}) apply  and $i=k$, then $g_\text{in}^\prime = g_\text{out}^\prime$, i.e., the metrics of both input and output vectors are equal.
    
    \begin{align}
        T = \hat{Y}_{11} S \in U(n,\hat{\eta}= \hat{\sigma}_3).
    \end{align}
    
    \begin{align}
         \begin{bmatrix} \psi_{01} \\ \psi_{11} \end{bmatrix}
         & = T 
        \begin{bmatrix} \psi_{00} \\ \psi_{10} \end{bmatrix}.
    \end{align}
        
    \begin{align}
        \begin{bmatrix} \psi_{01}^* & \psi_{11}^* \end{bmatrix}
        \begin{pmatrix} 1 & 0 \\ 0 & -1 \end{pmatrix}
        \begin{pmatrix} 1 & 0 \\ 0 & 1 \end{pmatrix}
        \begin{bmatrix} \psi_{01} \\ \psi_{11} \end{bmatrix}
        =
         \begin{bmatrix} \psi_{00}^* & \psi_{10}^* \end{bmatrix}
        \begin{pmatrix} 1 & 0 \\ 0 & -1 \end{pmatrix}
        \begin{pmatrix} 1 & 0 \\ 0 & 1 \end{pmatrix}
        \begin{bmatrix} \psi_{00} \\ \psi_{10} \end{bmatrix}.
    \end{align}

A subsequent application of $\hat{Y}_{01}$ leaves the dot-product conservation metrics unchanged, since $(g_\alpha^\prime (T))_{00} \neq (g_\alpha^\prime (T))_{11}$. The pseudo-unitary metric vector for $A$ is $\vec{ \hat{\alpha}}_3$, here shortened by the entry $\hat{\tau}_2$ of transposition around the anti-diagonal followed by product by $-1$.
    \begin{align}
        A = \hat{Y}_{01}T \in U \left(n, \hat{\eta}=-\hat{\tau}_2  \right).
    \end{align} 
        
    \begin{align}
         \begin{pmatrix} \psi_{10} \\ \psi_{11} \end{pmatrix}
        = A \begin{pmatrix} \psi_{00} \\ \psi_{01} \end{pmatrix}.
    \end{align}

    \begin{align}
         \begin{bmatrix} \psi_{10}^* & \psi_{11}^* \end{bmatrix}
        \begin{pmatrix}
           1 & 0
        \\ 0 & 1
        \end{pmatrix}
        \begin{pmatrix}
           1 & 0
        \\ 0 & -1
        \end{pmatrix}
        \begin{pmatrix}
           1 & 0
        \\ 0 & 1
        \end{pmatrix}
        \begin{bmatrix} \psi_{10} \\ \psi_{11} \end{bmatrix} 
         =
        \begin{bmatrix} \psi_{00}^* & \psi_{01}^* \end{bmatrix}
        \begin{pmatrix}
           1 & 0
        \\ 0 & 1
        \end{pmatrix}
        \begin{pmatrix}
           1 & 0
        \\ 0 & -1
        \end{pmatrix}
        \begin{pmatrix}
           1 & 0
        \\ 0 & 1
        \end{pmatrix}
        \begin{bmatrix} \psi_{00} \\ \psi_{01} \end{bmatrix}.
    \end{align}

An example in which both the input and output dot-product conservation metrics are different is given by the operator $B$, whose group-like object is fully defined by node $\vec{\hat{\phi}}_6$ in figure \ref{hex}.

    \begin{align} \label{aalaw0}
        B = \hat{Y}_{01}S \in U\left(n, \hat{\eta}=- \hat{\tau}_1 \right).
    \end{align}
    \begin{align}
         \begin{bmatrix} \psi_{11} \\ \psi_{10} \end{bmatrix}
        = B \begin{bmatrix} \psi_{00} \\ \psi_{01} \end{bmatrix}.
    \end{align}
    \begin{align}    
         \label{aalaw}
        \begin{bmatrix} \psi_{11}^* & \psi_{10}^* \end{bmatrix}
        \begin{pmatrix}
           -1 & 0
        \\ 0 & 1
        \end{pmatrix}
        \begin{bmatrix} \psi_{11} \\ \psi_{10} \end{bmatrix}
        =     
        \begin{bmatrix} \psi_{00}^* & \psi_{01}^* \end{bmatrix}
        \begin{pmatrix}
           1 & 0
        \\ 0 & -1
        \end{pmatrix}
        \begin{bmatrix} \psi_{00} \\ \psi_{01} \end{bmatrix} .
    \end{align}

Successive application of $B$ could lead to an ill-defined dot-product conservation metric for medial terms. This ambiguity is suppressed by alternating $B$ with its counterpart $\hat{Y}_{10}S$, whose vector space takes the same dot-product conservation metrics as $B$ with opposite signs. Both belong to the same group-like object $U(2, \vec{\hat{\phi}}_6) $ that squares to the group $U(2, \vec{\hat{\epsilon}}_6)$.

\section{\label{sec5}Applications} 
\subsection{Invertibility through singular matrices}
The fact that the matrix inversion algorithm through partial inversion inverts diagonal entries may pose a problem for matrices with zeros along the main diagonal, even if they are invertible. In this section, we propose a solution to this issue by manipulating zeros as variables in a special set defined here.
	We proceed by analyzing the conditions that make hypothesis $H$ true, with a finite matrix being just a matrix without infinities.

\vspace{1em}
\fbox{\begin{minipage}{14cm}
{$H$}: The partial inversion algorithm applied successively along the main diagonal of a matrix produces the finite inverse matrix for any finite invertible matrix.
\end{minipage}
}
\vspace{1em}

If hypothesis $H$ is true, then invertible matrices such as permutation matrices can lead us to define a formalism to manipulate zeros and infinities in order to obtain a finite matrix. We take as an example the simplest of permutation matrices, $\pi_1$, without simplifications other than distributivity and associativity, putting symbols $o$ in place of zeros (equations (\ref{elast0}) through (\ref{elast})).

\begin{align} \label{elast0}
\gamma =
\hat{Y}_{11}\hat{Y}_{00} \pi 
&= \hat{Y}_{11}\hat{Y}_{00}
\begin{pmatrix}
   o & 1
\\ 1 & o
\end{pmatrix}
\\ &= \hat{Y}_{11}
\begin{pmatrix}
   \frac{1}{o} & -\frac{1}{o}
\\ \frac{1}{o} & o - \frac{1}{o}
\end{pmatrix}
\\ &= \hat{Y}_{11}
\begin{pmatrix}
   \frac{1}{o} & -\frac{1}{o}
\\ \frac{1}{o} & \frac{o^2-1}{o}
\end{pmatrix}
\\ &=
\begin{pmatrix}
   \frac{1}{o} +\frac{1}{o^2} \left(\frac{o}{o^2-1} \right) & -\frac{1}{o} \left(\frac{o}{o^2-1} \right)
\\ -  \frac{1}{o} \left(\frac{o}{o^2-1} \right) & \frac{o}{o^2-1}
\end{pmatrix}.
\label{elast}
\end{align}

Let $o \in \mathbb{W}$ be the basis of the set $\mathbb{W}$, with $\mathbb{M} = \mathbb{R} \cup \mathbb{W}$ a set that extends the real set $\mathbb{R}$, and consider  equations (\ref{eqst66}) through (\ref{eqstt}), with $\{N,M,p\} \in \mathbb{R}, \quad p \neq 0$.
\begin{align} \label{eqst66}
N+Mo^p &\in \mathbb{M}; 
\\ Real(N+Mo^p) &= \lim_{x \to 0} (N+Mx^p) \in \mathbb{R};
\\ Mem(N+0 ) &= N+Mo^{p} \in \mathbb{M};
\\ mem(N+Mo^p) &= Mo^p \in \mathbb{W}.
\label{eqstt}
\end{align}

In $\mathbb{W}$, differently from the real set, zeros are conveniently quantifiable, or equivalently, in a more computational terminology, their occurrences consume memory. It follows that equations (\ref{quan}) through (\ref{quan2}) are true in $\mathbb{W}$.
\begin{align} \label{quan}
 o &\neq \sum^N_i o_i = No,
 \\ o &\neq \prod^N_i o_i = o^N,
 \\ o + N &\neq N.
 \label{quan2}
\end{align}

Since $Real(\pi_1) = Real(\pi_1)^{-1}$ in real space, then the realization of the inverse in $\mathbb{W}$ should preserve the equality if the hypothesis $H$ is true, i.e., $Real(Mem(\pi) ) = Real(Mem(\pi)^{-1} )$. With this assumption,  equation (\ref{elast}) implies  equations (\ref{impos0}) through (\ref{eow2}), with $\gamma_{ij}$ the entries of the matrix $\gamma$.
\begin{align}
Real(\gamma_{00} ) &=0, \label{impos0}
\\ Real(\gamma_{01} ) &=1, \label{impos1}
\\ Real(\gamma_{10} ) &=1, \label{impos2}
\\ Real(\gamma_{11} ) &=0.
\label{eow2}
\end{align}

In order to investigate possible ambiguities during simplification, let us assume the results $a,b,c,d$ of equations (\ref{don}) through (\ref{don2}) are unknown, so that we can seek their solutions in $\mathbb{M}$ that allow $H$ to be true.
\begin{align} \label{don}
\frac{o}{o} &= a,
\\ 1-1 &=b,
\\ o-o &=c,
\\ \frac{1}{o} - \frac{1}{o} &= d.
\label{don2}
\end{align}

For $\gamma_{11}$, notice that $Real(\gamma_{11} ) = \frac{0}{0-1} =0$. For $\gamma_{01}$ and $\gamma_{10}$, one can put the numerator $o$ in evidence and apply $o/o = a$.

\begin{align}
-\frac{1}{o} \left(\frac{o}{o^2 - 1} \right) &=
-\frac{o}{o} \left(\frac{1}{o^2 - 1} \right)
\\ &= \frac{-a}{o^2 - 1}.
\label{era}
\end{align}

Upon realization in equation (\ref{tho}), and considering constraints defined in equations (\ref{impos1}) and (\ref{impos2}), equation (\ref{era}) is true iff $a=1$.
\begin{align} \label{tho}
Real\left(\frac{-a}{o^2 - 1} \right) = \frac{-a}{0 - 1} = \frac{a}{1}.
\end{align}

For $\gamma_{00}$, and considering we already found $o/o=1$, let us evidence $\frac{1}{o}$ for the whole expression, put the numerator $o$ in evidence and simplify the sum of fractions into one fraction, in equations (\ref{gamma00s}) through (\ref{denow}).

\begin{align} \label{gamma00s}
\gamma_{00} = \frac{1}{o} + \frac{1}{o^2} \left(\frac{o}{o^2-1} \right)
&=\frac{1}{o} \left[1 + \frac{1}{o} \left(\frac{o}{o^2-1} \right) \right]
\\ &=\frac{1}{o} \left[1 + \frac{o}{o} \left(\frac{1}{o^2-1} \right) \right]
\\ &= \frac{1}{o} \left[1 +  \frac{1}{o^2-1} \right]
\\ &= \frac{1}{o} \left[\frac{ o^2-1 + 1}{o^2-1} \right]
\\ &= \frac{ o^2-1 + 1}{o^3-o} .
\label{denow}
\end{align}

Dividing both the numerator and denominator of equation (\ref{denow}) by $o$ (equations (\ref{eyt}) through (\ref{eyt4})), the denominator becomes realizable to $-1$, so the constraint defined in equation (\ref{impos0}) implies the numerator must be realizable to zero, leading to equation (\ref{eyt4}).
\begin{align} \label{eyt}
\gamma_{00} &= \frac{ o -\frac{1}{o} + \frac{1}{o} }{o^2-1}
\\ &= \frac{ o +\frac{1}{o} (1-1) }{o^2-1}
\\ &= \frac{ o +\frac{1}{o} b }{o^2-1} .
\label{eyt3}
\\ Real\left(o +\frac{b}{o}  \right) &=0 \quad \iff \quad Real\left(\frac{b}{o} \right) = 0.
\label{eyt4}
\end{align}

Finally, $Real(\frac{b}{o} ) = 0$ requires $b = Mo^n$, with $M,n$ real numbers, $M\neq 0$, and $n>1$ .

Once $1-1$ is defined, we can multiply both sides by $o$ and find the values of $c$ and $d$.
\begin{align}
c = o(1-1) &= o Mo^n
\\ \notag =o-o &= Mo^{n+1}.
\end{align}
and
\begin{align}
d= \frac{1}{o}(1-1) &= \frac{1}{o} Mo^n
\\  \notag =\frac{1}{o}- \frac{1}{o} &= Mo^{n-1}.
\end{align}

Repeating the previous steps for other powers, we obtain equation (\ref{genr}). The conditions on $n$ to be a sufficiently large positive is to make $Mo^{n+p}$ realizable to $0$ for any power $p$, including negatives and $p=0$, but excepting $p=-n$, to be discussed later.
\begin{align} \label{genr}
o^p - o^p &= Mo^{n+p}, 
\\ \notag \text{with} \quad &n >|p|, \quad  n>0, \quad \{n,p\} \in \mathbb{R}.
\end{align}

It may be intuitive to try to conceive $1-1 = o$ as true, however, it implies $o^{-1} - o^{-1} = 1$, and recurrently leads to $o^p - o^p = o^{p+1}$ and to $o=2$, which is obviously false under realization. Therefore the definition of the exponent $n$ makes the algorithm realizable without inconsistencies.

The reversibility of this model can be attested even in non-invertible matrices (e.g., equations (\ref{ewq}) through (\ref{mems}), using the entry-wise algorithm defined in equations (\ref{e:pialgo}) through (\ref{e:pialgo2})). In this case, non-invertibility of $M$ means that no matrix can multiply $M$ and result in an identity matrix, even though the partial inversion algorithm can be used to invert it reversibly in $\mathbb{W}$.

\begin{align} \label{ewq}
X &= \begin{pmatrix}
o & o
\\ o & o
\end{pmatrix}
\\
\hat{Y}_{11}\hat{Y}_{00}\hat{Y}_{11}\hat{Y}_{00} X
&= \hat{Y}_{11}\hat{Y}_{00}\hat{Y}_{11} \begin{pmatrix}
o^{-1} & -1
\\ 1 & o - \frac{o^2}{o}
\end{pmatrix}
\\ &= \hat{Y}_{11}\hat{Y}_{00}\hat{Y}_{11} \begin{pmatrix}
o^{-1} & -1
\\ 1 & o - o
\end{pmatrix}
\\
&= \hat{Y}_{11}\hat{Y}_{00} \begin{pmatrix}
	\frac{1}{o} + \frac{1}{o-o}        & -\frac{1}{o-o}
\\ -\frac{1}{o-o} & \frac{1}{o-o}
\end{pmatrix}
\\ &= \hat{Y}_{11}\hat{Y}_{00}
\begin{pmatrix}
   \frac{1-1+1 }{Mo^{n+1}} &  \frac{-1 }{Mo^{n+1}}
\\ \frac{-1 }{Mo^{n+1}} & \frac{1 }{Mo^{n+1}}
\end{pmatrix}.
\label{mems}
\end{align}

Use of $1-1=Mo^n$ implies two simplification routes for $1-1+1$, that give different results in the memory set, but the same result in $\mathbb{R}$. By considering $(1-1)+1 = Mo^n+ 1$ at the last equation of  \ref{mems}:
\begin{align}
\hat{Y}_{11}&\hat{Y}_{00} 
\begin{pmatrix}
   \frac{1-1+1 }{Mo^{n+1}} &  \frac{-1 }{Mo^{n+1}}
\\ \frac{-1 }{Mo^{n+1}} & \frac{1 }{Mo^{n+1}}
\end{pmatrix}
= \hat{Y}_{11}\hat{Y}_{00}
\begin{pmatrix}
   \frac{Mo^n + 1 }{Mo^{n+1}} &  \frac{-1 }{Mo^{n+1}}
\\ \frac{-1 }{Mo^{n+1}} & \frac{1 }{Mo^{n+1}}
\end{pmatrix}
\\ &= \hat{Y}_{11}
\begin{pmatrix}
\frac{Mo^{n+1}}{Mo^n +1} & \frac{1}{Mo^{n} +1}
\\  \frac{-1}{Mo^{n} +1} & \frac{1}{Mo^{n+1}} - \frac{1  }{ (Mo^{n+1})^2 }
\frac{Mo^{n+1} }{Mo^{n} + 1}
\end{pmatrix}
\\ &=\hat{Y}_{11}
\begin{pmatrix}
\frac{Mo^{n+1}}{Mo^n +1} & \frac{1}{Mo^{n} +1}
\\  \frac{-1}{Mo^{n} +1} & \frac{ Mo^n + 1 - 1  }{ (Mo^{n+1}) (Mo^n + 1) }
\end{pmatrix}
\\ &= \hat{Y}_{11}
\begin{pmatrix}
\frac{Mo^{n+1}}{Mo^n +1} & \frac{1}{Mo^{n} +1}
\\  \frac{-1}{Mo^{n} +1} & \frac{ 2}{ o(Mo^n + 1) }
\end{pmatrix}
\end{align}
\begin{align}
 &=
\begin{pmatrix}
\frac{Mo^{n+1}}{Mo^n +1} + \frac{1}{(Mo^{n} +1)^2} \frac{o(Mo^n + 1) }{ 2 }        & \frac{o(Mo^n + 1) }{ 2 }  \frac{1}{Mo^{n} +1}
\\ \frac{1}{Mo^{n} +1}\frac{o(Mo^n + 1) }{ 2 }  & \frac{o(Mo^n + 1) }{ 2 }
\end{pmatrix}
\\ &=
\begin{pmatrix}
\frac{Mo^{n+1} +o  }{2 (Mo^n +1)}  & \frac{o}{2}
\\ \frac{o}{2} & \frac{Mo^{n+1} + o }{ 2 }
\end{pmatrix}. \label{finale}
\end{align}

Since elements $X$ in $\mathbb{W}$ retain information about occurrence of some operations, $Mem(X) \neq (Mem(X)^{-1})^{-1}$, however the equation $Real(Mem(X)) = Real((Mem(X)^{-1})^{-1} )$ even for non-invertible matrices, as can be seen from the last equation in equations (\ref{finale}).

Starting from the last equation of  equations (\ref{mems}) by considering the alternative simplification $(1+1)-1=1$, the realization of the result is consistently the realization of the initial matrix (equations (\ref{reaz}) through (\ref{reaz3})).
\begin{align} \label{reaz}
\hat{Y}_{11} \hat{Y}_{00} &
\begin{pmatrix}
   \frac{1-1+1 }{Mo^{n+1}} &  \frac{-1 }{Mo^{n+1}}
\\ \frac{-1 }{Mo^{n+1}} & \frac{1 }{Mo^{n+1}}
\end{pmatrix}
= \hat{Y}_{11}\hat{Y}_{00}
\begin{pmatrix}
  \frac{1}{Mo^{n+1}} &   \frac{-1}{Mo^{n+1}}
\\   \frac{-1}{Mo^{n+1}}  &   \frac{1}{Mo^{n+1}}
\end{pmatrix}
\\&= \hat{Y}_{11}
\begin{pmatrix}
   Mo^{n+1} &   1
\\ -1  &   \frac{1}{Mo^{n+1}} - \frac{1}{Mo^{n+1} }
\end{pmatrix}
\\ &= \hat{Y}_{11}
\begin{pmatrix}
   Mo^{n+1} &   1
\\ -1  &   \frac{Mo^{n}}{Mo^{n+1}  }
\end{pmatrix}
 \\&= \hat{Y}_{11}
\begin{pmatrix}
   Mo^{n+1} &   1
\\ -1  &   \frac{1}{o}  
\end{pmatrix}
 \\ &=
\begin{pmatrix}
   Mo^{n+1} + o &   o
\\ o  &   o
\end{pmatrix}.
\label{reaz3}
\end{align}

The actual values and nature of $M,n$ were vaguely defined earlier. In addition to  equations (\ref{don}) throught (\ref{don2}),  equations (\ref{totalr}) through (\ref{totalr2})  give the last axiom to sustain the set $\mathbb{W}$, the axiom of total realizability. Differently from the treatment we gave up to now, it starts with proposing that each application of $1-1=Mo^n$ results in an independent and unique pair of variables $M,n$, i.e., the generation of  $Mo^n \in \mathbb{W}$ is \textit{not necessarily} copiable.

\begin{align} \label{totalr}
Real(N) &= Real(N),
\\ Real(N) &= Real\left(N + \sum^P_{i=0} (1-1)_i \right),
\\ Real(N) &= Real\left(N +   \sum_{i=0}^P M_io^{n_{i}} \right),
\\ Real(N-N) &= Real(N(1-1)) 
\\ \notag &= Real(N(M_t o^{n_t})) 
\\ \notag &=Real\left( \sum_{i=0}^P M_io^{n_{i}} \right),
\end{align}

\begin{align}
 Real(N) = Real\left(\sum_{i=0}^P  \frac{M_i}{M_t} o^{n_{i}-n_t} \right).
\label{totalr2}
\end{align}

 Equations (\ref{totalr}) through (\ref{totalr2}) pose the problem that either $Mo^n$ is ill-defined and inconsistent with $o$-number-processing,  or $M,n$ are variables that are constrained by real equations upon application of the realization function $Real$.

The definition of $o$ and the series of statements that support it in this paper are meant to allow  for the simplification of expressions in the partial inversion algorithm  when it requires a bulky manipulation of zero-expressions, and thus they were, from the start, required to be in agreement with the equivalent equations in the real set. In other words, the proposed definitions may be conveniently defined as true in a computer program to serve their simplification purpose, although their consistency with set-theoretical definitions was not confronted.

Having the memory set $\mathbb{W}$ well defined, partial inversion allows us to study singular matrices and seek finite representations of them.

\subsection{Renormalized-growth approach to renormalization}
The S-matrix theory interpretation of pseudo-unitary groups allows us to look at partial inversion as a means to change the direction of the graph flow. This is particularly important in growing diagrams using operators that grow a diagram perpendicularly. This is the case of Chalker and Coddington's $M$ and $M^\prime$ \cite{chalker1988percolation}. In equation (\ref{erq}) we defined a matrix $A=\hat{Y}_{01}\hat{Y}_{11}S$ (in place of $M^\prime$) that grows a scattering diagram perpendicularly to a transfer matrix $T$ by matrix product. We use the subscript notation $_{(a,b)}$ to indicate the dimensions of the circuit represented by the matrix.

\begin{align} \label{erq}
    T_{(4,4)} = (\hat{Y}_{01} [A_{(2,2)}^3])^3.
\end{align}

\begin{figure*}
\caption{Equality of two renormalized-growth paths shows agreement with graphical interpretation of scattering lattice. \label{tworen} }
\begin{tabular}{c}
\includegraphics[width=0.45\linewidth]{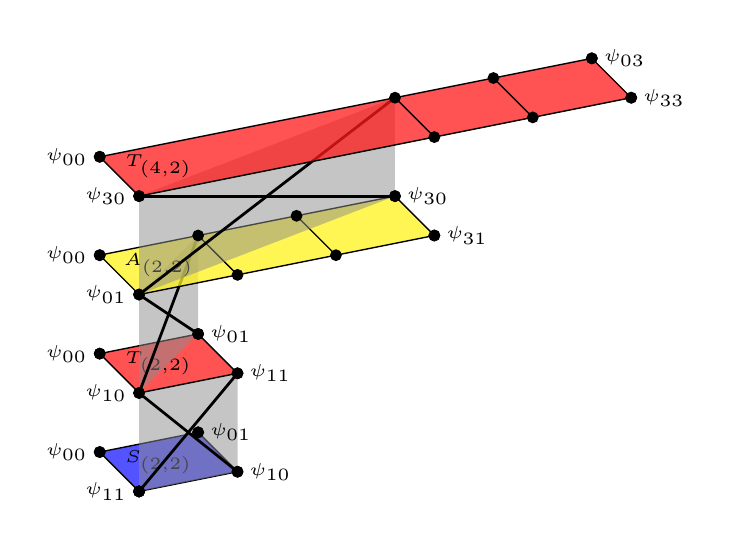} \\
\end{tabular}
  $=$ 
\begin{tabular}{c}
 \includegraphics[width=0.45\linewidth]{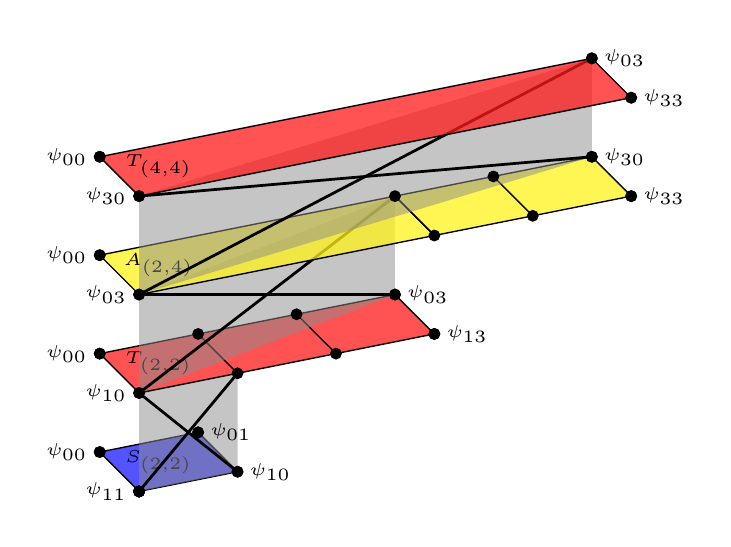} \\
\end{tabular} 
$.$
\end{figure*}

The more general case, for any positive integers $n,m$, is given in equation (\ref{equivrel}), and illustrated in figure \ref{tworen}.
\begin{align}
\label{equivrel}
    T_{(2n,m+1)} &= (\hat{Y}_{01} [(\hat{Y}_{01}T)^{2n-1} ])^m
    \\ \notag &=  \hat{Y}_{01}[(\hat{Y}_{01} [T^{m} ])^{2n-1}]
    .
\end{align}
The requirement of the odd number $2n-1$ is due to the fact that $A^{2n}$ is diagonal. Therefore, a partial inversion at indices $(0,1)$ would invert a zero, leading to a matrix with infinities. It is treatable using the special set of simplification rules defined in the previous section, but it was avoided in this example for simplicity.

The renormalized matrix for a diagram of dimensions $(4,4)$ is equation (\ref{kop}). It leads reasonably to the renormalization equations (\ref{reneq}) through (\ref{reneq4}).

\begin{align} \label{kop}
    T_{(4,4)} = 
    \begin{bmatrix}
         \frac{3r^2(r^*)^2 + rr^* }{(2rr^* -1)(t^*)^{3}} 
         & \frac{[r^2 (r^*)^2 + tt^* +2   ]r^3}{(2rr^* -1)(t^*)^{3}}
        \\ \frac{[r^2 (r^*)^2 + tt^* +2   ](r^*)^3}{(2rr^* -1)(t^*)^{3}} 
        & \frac{3r^2(r^*)^2 - 3rr^*}{(2rr^* -1)(t^*)^{3}}
    \end{bmatrix}
\end{align}

\begin{align}
    \label{reneq}
    T^{--} &= \frac{3r^2(r^*)^2 - 3rr^* }{(2rr^* -1)(t^*)^{3}} ,
    \\ \label{reneq2} T^{++} &= \frac{3r^2(r^*)^2 + rr^* }{(2rr^* -1)(t^*)^{3}},
    \\ \label{reneq3} T^{+-}   &=  
    \frac{(r^2 (r^*)^2 + tt^* +2   )r^3}{(2rr^* -1)(t^*)^{3}}
    \\ T^{-+} &=
    \frac{(r^2 (r^*)^2 + tt^* +2   )(r^*)^3}{(2rr^* -1)(t^*)^{3}}
    \label{reneq4}
\end{align}

The graphical interpretation of Scattering Theory gives the partial inversion representation a physical meaning by changing the diagrammatic time flow from linear (in S-matrix formalism) to circular (in T-matrix formalism), and any other configuration in between. The diagrammatic time flow is accompanied by the charge ordering of the scattering process, with space-time coordinates and other properties of the wavefunctions as generalized charges. By choosing a preferential direction for diagrammatic time flow, i.e., a partial inversion of the formalism, one chooses for which charge the matrix of the formalism is a translation operator of, with the S-matrix as translation operator of time levels, and T-matrix as translation operator of space levels.

The restrictions imposed on unitary systems by No-Go Theorems (no-cloning, no-deleting, no-masking) are largely studied in pseudo-unitary systems due to the possibility of flexibilizing these theorems with a set of $\eta$-orthogonal states \cite{zhan2020experimental, lv2022quantum}. 
These models focus on pseudo-unitary systems alone without realizing the transition from unitary systems to $\eta$-pseudo-unitary ones and vice-versa.
Since $\eta$-orthogonal cloning was proved possible \cite{zhan2020experimental}, in order to exploit it in unitary systems, a pair of desunitarization and reunitarization protocols is necessary. 

The closure property of Lie groups does not allow a unitary gate to generate a different $\eta$-pseudo-unitary one. This implies that no observable is related to the conversion between $\eta$-pseudo-unitary groups, and a gate representation is not required for the conversion. 
Instead, the operation of partial inversion as defined in this paper allows for the interpretation of the conversion in an abstract way, so that an $\eta$-pseudo-unitary circuit can exist as an abstraction of a unitary circuit. In particular, we can create a circuit such that it starts as unitary, gets converted into an $\eta$-pseudo-unitary circuit, then extended with $\eta$-pseudo-unitary gates, and finally converted back into a unitary circuit. After the restoration of unitarity, the $\eta$-pseudo-unitary part becomes the application of a parameterized general unitary gate. 

A possible advantage of using an $\eta$-pseudo-unitary circuit in place of a unitary one is purely syntactic, by building a $\sigma_3$-pseudo-unitary circuit from compositions of $W$ in equation (\ref{zzpauli}), and then using partial inversion to obtain its unitary representation. This means that with an appropriate desunitarization-reunitarization protocol in terms of partial inversion, an $\eta$-pseudo-unitary circuit is an alternative way of modelling unitary circuits.

\begin{figure}[h]
    \centering
    \caption{Landauer conductance of real T-matrix for diagrammatic dimensions from $(2,2)$ to $(10,10)$ after application of the renormalized-growth algorithm.}
    \includegraphics[width=\linewidth]{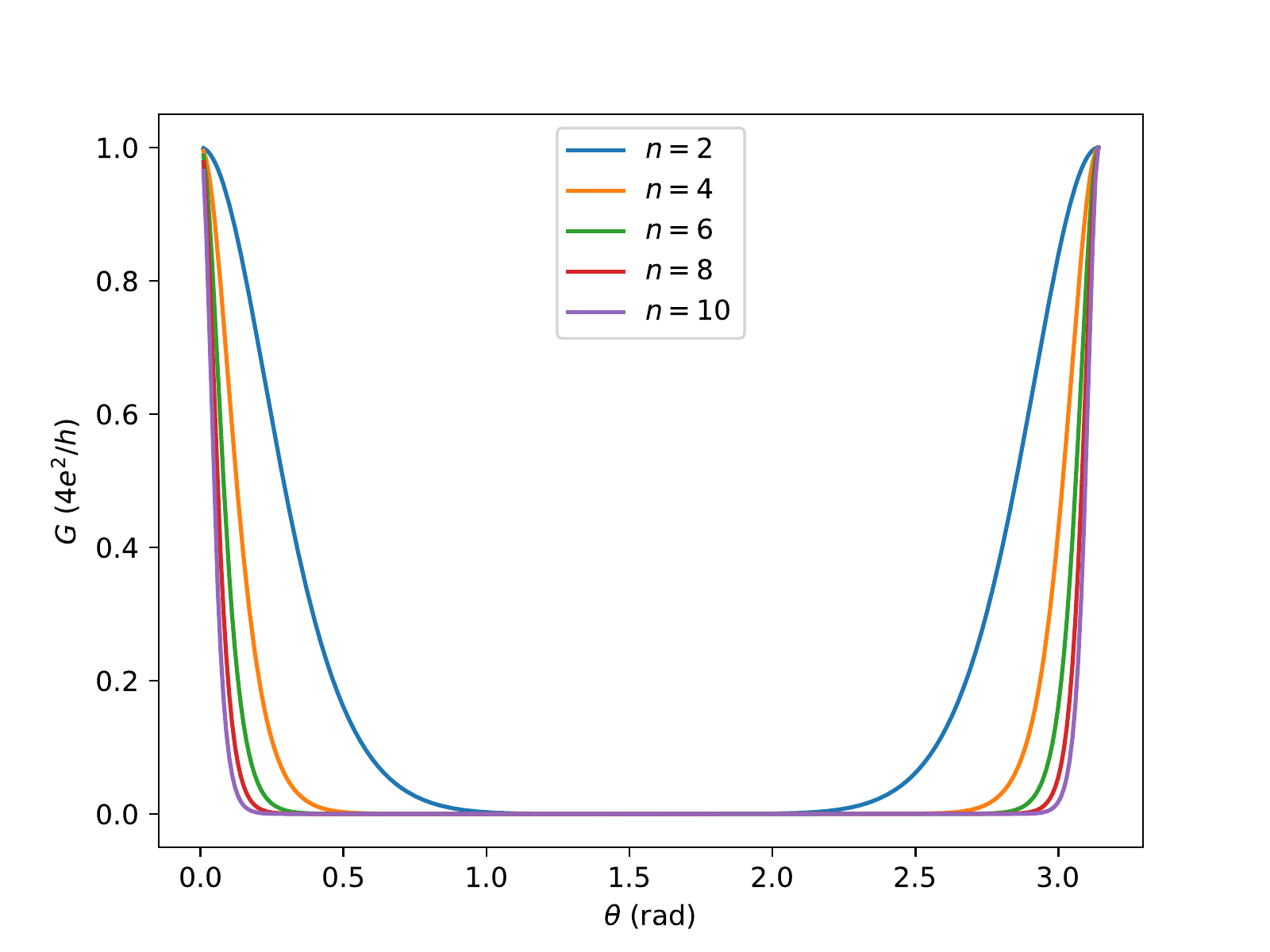}
    \label{figland}
\end{figure}

In the light of the Tensor Renormalization Group, the graphical process of partial inversion is regarded as rewiring \cite{cook2015tensor}. As was shown in the Renormalized-Growth algorithm, a special sequence of partial inversions and matrix powers allows for the manipulation of corner states of a scattering lattice. This algorithm is an example of a Tensor Renormalization procedure that makes constant use of rewiring to re-scale the circuit for use in lower dimensions. The experimental implications of this algorithm can be estimated by calculation of Landauer conductance. Figure \ref{figland}, with $t=\cos(\theta)$ and $r=\sin(\theta)$ shows that increasing the dimensions of the scattering diagram increases the scattering and reduces the transmission along different values of $\theta$.

\section{\label{sec6}Conclusion}
In this paper, we proposed a matrix formalism for the transformation in matrix space that results from permuting input and output vector entries. We call this operation partial inversion, since it can be used to invert matrices step-wise. As we discussed, partial inversion affects the dot product in a similar way to the Legendre transform that served as inspiration, with the difference that, in partial inversion, a metric is defined such that the new dot product is always a conserved quantity, regardless of its physical significance.

We used permutation planes in a 3D grid as diagram representation of partial inversion in a quantum circuit, and defined some of its properties. This representation eases the further illustration of inter-pseudo-unitary circuits, i.e., circuits that transit between $\eta$-pseudo-unitary groups.

We defined the rules that allow for the application of partial inversion even on entries with zeros by defining a special set $\mathbb{M}$ that treats zero as a numerical basis.

S-matrix theory provides us with examples to analyze metric changes in pseudo-unitarity groups due to partial inversion, as it is already used to map S-matrices to T-matrices; however without a proper name other than solving a system of equations.
We explored the fact that this transformation changes the geometric growth of the circuit to propose a renormalized-growth algorithm. 
For example, if a matrix $T$ grows a circuit along the $x$ axis, by matrix product, then we can use partial inversion to find the $A$ matrix that will continue the growth by matrix product along a perpendicular $y$ axis. The final matrix is a representation of a larger two-dimensional circuit without changing the matrix dimensions.

We found the $(2,2)$-dimensional $T$ matrix that represents a $(4,4)$-dimensional system symbolically as a test of the proposed renormalized growth. Our results can potentially be used to devise other inter-unitary circuits that can be conservatively transformed into unitary circuits, expanding the possibilities for quantum computing algorithms.

\section{\label{acknowledgements}Acknowledgements} 
The first author is thankful to the fruitful discussions with prof. Dmitry Melnikov at the International Institute of Physics (IIP) in the Federal University of Rio Grande do Norte (UFRN), Brazil. This article was partially funded by the Coordenação de Aperfeiçoamento de Pessoal de Nível Superior - Brasil (CAPES) via affiliation of the first author to the Graduate Physics Program of UFRN, and by QC2.

\section{Data availability statement}
The data that support the findings of this study are available upon request from the authors.


\bibliographystyle{unsrt}

\clearpage
\end{document}